\documentclass[twocolumn]{aastex63}

\usepackage{graphicx}
\usepackage{amsmath}
\usepackage{txfonts}
\usepackage{xcolor}
\usepackage{url}
\usepackage{xspace}

\newcommand{\muHz}{\,\mu\mathrm{Hz}}
\newcommand{\K}{\,\mathrm{K}}
\newcommand{\Msun}{\,\mathrm{M}_\odot}

\newcommand{\numax}{\nu_{\mathrm{max}}\xspace}
\newcommand{\dnu}{\Delta\nu\xspace}
\newcommand{\dotwo}{\delta\nu_{02}}
\newcommand{\Teff}{\,T_\mathrm{eff}}
\newcommand{\eps}{\,\varepsilon}
\newcommand{\bprp}{\,G_{\mathrm{BP}}-G_{\mathrm{RP}}}

\newcommand{\kicexample}{KIC4448777\xspace}
\newcommand{\priornumber}{$13\,288$\xspace}
\newcommand{\kde}{\texttt{KDE}\xspace}
\newcommand{\asypeakbag}{\texttt{Asy\_peakbag}\xspace}
\newcommand{\peakbag}{\texttt{Peakbag}\xspace}
\newcommand{\emcee}{\texttt{emcee}\xspace}
\newcommand{\cpnest}{\texttt{CPNest}\xspace}

\newcommand{\pymc}{\texttt{PyMC3}\xspace}
\newcommand{\pbjam}{\texttt{PBjam}\xspace}
\newcommand{\pmodes}{p-modes\xspace}
\newcommand{\pmode}{p-mode\xspace}
\newcommand{\gmodes}{g-modes\xspace}

\newcommand{\kepler}{\textit{Kepler}\xspace}
\newcommand{\tess}{TESS\xspace}
\newcommand{\corot}{CoRoT\xspace}
\newcommand{\plato}{PLATO\xspace}

\usepackage{CJKutf8}
\newcommand{\JOchinesename}{{\begin{CJK}{UTF8}{gbsn}(王加冕)\end{CJK}}}
\newcommand{\TLchinesename}{{\begin{CJK}{UTF8}{gbsn}(李坦达)\end{CJK}}}

\received{July 9, 2020}
\accepted{November 22, 2020}
\submitjournal{AJ}
\shorttitle{PBjam}
\shortauthors{Nielsen et al.}
\graphicspath{{./}}

\begin{document}

\title{PBjam\footnote{Release 1.0.0. \url{https://doi.org/10.5281/zenodo.4300079}}: \\A Python package for automating asteroseismology of solar-like oscillators}

\correspondingauthor{M. B. Nielsen}
\email{m.b.nielsen.1@bham.ac.uk}

\author[0000-0001-9169-2599]{M. B. Nielsen}
\affiliation{School of Physics and Astronomy, University of Birmingham, Birmingham B15 2TT, UK}
\affiliation{Stellar Astrophysics Centre (SAC), Department of Physics and Astronomy, Aarhus University, Ny
Munkegade 120, DK-8000 Aarhus C, Denmark}
\affiliation{Center for Space Science, NYUAD Institute, New York University Abu Dhabi, PO Box 129188, Abu
Dhabi, United Arab Emirates}

\author[0000-0002-4290-7351]{G. R. Davies}
\affiliation{School of Physics and Astronomy, University of Birmingham, Birmingham B15 2TT, UK}

\author[0000-0002-4773-1017]{W. H. Ball}
\affiliation{School of Physics and Astronomy, University of Birmingham, Birmingham B15 2TT, UK}
\affiliation{Stellar Astrophysics Centre (SAC), Department of Physics and Astronomy, Aarhus University, Ny
Munkegade 120, DK-8000 Aarhus C, Denmark}

\author[0000-0001-8355-8082]{A. J. Lyttle}
\affiliation{School of Physics and Astronomy, University of Birmingham, Birmingham B15 2TT, UK}
\affiliation{Stellar Astrophysics Centre (SAC), Department of Physics and Astronomy, Aarhus University, Ny
Munkegade 120, DK-8000 Aarhus C, Denmark}

\author[0000-0001-6396-2563]{T. Li \TLchinesename}
\affiliation{School of Physics and Astronomy, University of Birmingham, Birmingham B15 2TT, UK}
\affiliation{Stellar Astrophysics Centre (SAC), Department of Physics and Astronomy, Aarhus University, Ny
Munkegade 120, DK-8000 Aarhus C, Denmark}

\author[0000-0002-0468-4775]{O. J. Hall}
\affiliation{School of Physics and Astronomy, University of Birmingham, Birmingham B15 2TT, UK}
\affiliation{Stellar Astrophysics Centre (SAC), Department of Physics and Astronomy, Aarhus University, Ny
Munkegade 120, DK-8000 Aarhus C, Denmark}

\author[0000-0002-5714-8618]{W. J. Chaplin}
\affiliation{School of Physics and Astronomy, University of Birmingham, Birmingham B15 2TT, UK}
\affiliation{Stellar Astrophysics Centre (SAC), Department of Physics and Astronomy, Aarhus University, Ny
Munkegade 120, DK-8000 Aarhus C, Denmark}

\author[0000-0001-8330-5464]{P. Gaulme}
\affiliation{Max-Planck-Institut f\"ur Sonnensystemforschung, Justus-von-Liebig-Weg 3, 37077, G\"ottingen}

\author[0000-0003-1001-5137]{L. Carboneau}
\affiliation{School of Physics and Astronomy, University of Birmingham, Birmingham B15 2TT, UK}
\affiliation{Stellar Astrophysics Centre (SAC), Department of Physics and Astronomy, Aarhus University, Ny
Munkegade 120, DK-8000 Aarhus C, Denmark}

\author[0000-0001-7664-648X]{J. M. J. Ong \JOchinesename}
\affiliation{Department of Astronomy, Yale University, 52 Hillhouse Ave., New Haven, CT 06511, USA}

\author[0000-0002-8854-3776]{R. A. García}
\affiliation{IRFU, CEA, Universit\'e Paris-Saclay, F-91191 Gif-sur-Yvette, France}
\affiliation{AIM, CEA, CNRS, Universit\'e Paris-Saclay, Universit\'e Paris Diderot, Sorbonne Paris Cit\'e, F-91191 Gif-sur-Yvette, France}

\author[0000-0002-7547-1208]{B. Mosser}
\affiliation{LESIA, Observatoire de Paris, Universit\'e PSL, CNRS, Sorbonne Universit\'e, Universit\'e de Paris, 92195 Meudon, France}

\author[0000-0002-7403-2764]{I. W. Roxburgh}
\affiliation{Astronomy Unit, School of Physics and Astronomy, Queen Mary University of London, London E1 4NS, UK}
\affiliation{School of Physics and Astronomy, University of Birmingham, Birmingham B15 2TT, UK}

\author[0000-0001-8835-2075]{E. Corsaro}
\affiliation{INAF – Osservatorio Astrofisico di Catania, via S. Sofia 78, 95123 Catania, Italy}

\author[0000-0001-9405-5552]{O. Benomar}
\affiliation{Solar Science Observatory, NAOJ and Department of Astronomical Science, Sokendai (GUAS), Mitaka, Tokyo, Japan}
\affiliation{Center for Space Science, NYUAD Institute, New York University Abu Dhabi, PO Box 129188, Abu
Dhabi, United Arab Emirates}

\author[0000-0003-1665-5389]{A. Moya}
\affiliation{Electrical Engineering, Electronics, Automation and Applied Physics Department, E.T.S.I.D.I, Polytechnic University of Madrid (UPM), Madrid 28012, Spain}
\affiliation{School of Physics and Astronomy, University of Birmingham, Birmingham B15 2TT, UK}

\author[0000-0001-9214-5642]{M. N. Lund}
\affiliation{Stellar Astrophysics Centre (SAC), Department of Physics and Astronomy, Aarhus University, Ny
Munkegade 120, DK-8000 Aarhus C, Denmark}

\begin{abstract}
Asteroseismology is an exceptional tool for studying stars by using the properties of observed modes of oscillation. So far the process of performing an asteroseismic analysis of a star has remained somewhat esoteric and inaccessible to non-experts. In this software paper we describe \pbjam, an open-source Python package for analyzing the frequency spectra of solar-like oscillators in a simple but principled and automated way. The aim of \pbjam is to provide a set of easy-to-use tools to extract information about the radial and quadropole oscillations in stars that oscillate like the Sun, which may then be used to infer bulk properties such as stellar mass, radius and age or even structure. Asteroseismology and its data analysis methods are becoming increasingly important as space-based photometric observatories are producing a wealth of new data, allowing asteroseismology to be applied in a wide range of contexts such as exoplanet, stellar structure and evolution, and Galactic population studies.
\end{abstract}


\section{Introduction}
In the past few decades asteroseismology\footnote{Helioseismology in the case of the Sun.} has become an important tool for characterizing stars. The frequencies of modes in which many different stars oscillate are sensitive probes of their physical properties \citep{Aerts2010, Chaplin2013}. These mode frequencies are used as observational constraints in stellar modeling \citep[see, e.g.,][ for recent examples from \kepler]{Appourchaux2012, Davies2016a, Lund2017}, where they routinely allow for estimates of the physical properties of stars down to the percent level \citep[see, e.g.,][]{Metcalfe2012, Lebreton2014, SilvaAguirre2017}. The types of oscillations that a star can exhibit vary widely depending on its physical properties \citep{Handler2013}, but in the following we will focus on stars that oscillate like the Sun, namely those with convective layers at the surface. 

Asteroseismology uses time series from two main sources: either radial velocity or photometric intensity. These observations stem from what is now a multitude of different ground- and space-based observatories and projects, including the ground based GONG \citep{Harvey1996}, BiSON \citep{Hale2016} and SONG \citep{Grundahl2017} projects, and the SOHO \citep{Domingo1995}, \corot \citep{Baglin2009}, \kepler \citep{Borucki2010} and \tess \citep{Ricker2014} spacecraft. Analysis of the observations from this multitude of telescopes is in principle possible in the time domain \citep{celerite}, but this remains computationally expensive. Asteroseismology of stars that oscillate like the Sun is therefore typically performed in frequency space, where the oscillations modes are often readily visible.

In cool main-sequence (MS) stars like the Sun, as well as in sub-giant (SG) and red giant (RG) stars, the oscillations that are most clearly visible are acoustic \pmodes \citep{Garcia2019}. These are stochastically excited, damped harmonic oscillations, which take the form of a set of semi-regularly spaced near-Lorentzian peaks in the power spectrum \citep{Tassoul1980, Anderson1990}. The central frequencies of these peaks correspond to the resonance frequencies of the star, which can be compared to the frequencies of stellar structure models and so allow constraints to be placed on the properties of the star \citep[e.g.][]{Brown1994, SilvaAguirre2015, Angelou2017}. 

The measurement of the oscillation mode frequencies, also known as ``peakbagging", is the overall aim of \pbjam\footnote{Documentation and usage instructions are available at \url{https://pbjam.readthedocs.io}}. This is typically done by fitting a parametric model to the power spectrum of the photometric time series, for example, for each individual star \citep{Appourchaux2003b, Davies2014b, Corsaro2014}. This process involves two main parts: the mode identification and the model fitting, where the mode identification step informs the choice of the parametric model of the spectrum that is then fit to to the observed oscillation spectrum. 

Identifying the oscillation frequencies is often done manually and requires some knowledge of how the modes may appear at a given evolutionary stage of the star. The modes in most oscillating stars are described using spherical harmonic functions of angular degree $l$ and azimuthal order $m$, each of which have a number of overtones of radial order, $n$. The mode identification requires assigning unique labels $(n,l,m)$ to the modes that are visible in the spectrum. This is particularly difficult in SG and RG stars where the modes start coupling to the internal gravity modes \citep[\gmodes, see, e.g.,][]{Mosser2012b}, and thereby start rapidly varying in frequency as the star evolves. For F-type stars the mode widths increase substantially compared to cooler stars \citep{Appourchaux2014}, blending the $l=2,0$ modes in the spectrum and making them appear almost identical to the $l=1$ modes, thus any distinction is difficult \citep{Appourchaux2008, Benomar2009, White2012}. Both these cases as well as simpler ones, like Sun-like stars, also become further complicated when the signal-to-noise ratio (SNR) of the observations is low. 

Following the mode identification, the next step is fitting the chosen model to the spectrum of the oscillation frequencies. Finding the model that has the highest probability of explaining the observed spectrum can often be computationally expensive. This is especially true for observations made by the \kepler spacecraft, which observed more than $20,000$ oscillating stars for several years with a cadence of $\approx30$ minutes \citep[see, e.g.,][]{Yu2018}, and several hundred with a shorter cadence of $\approx 1$ minute \citep{Lund2017, Serenelli2017}. The spectra of many of these stars still need to be analyzed in detail. Furthermore, at the time of writing \tess is continuously producing new time series of stars across almost the entire sky, and with the future launch of the \plato \citep{Rauer2014} mission, a fast and automated method of peakbagging is becoming increasingly important.  

With \pbjam we focus on solving these two issues for solar-like oscillators in an automated fashion. These stars are the most numerous type of oscillator, and include MS stars with masses $\lesssim1.6\Msun$ and the vast majority of SG and RG stars. The former are of particular interest in, e.g., exoplanet studies \citep{Lundkvist2016, Huber2019}, and the latter for probing Galactic populations due to their visibility at great distances \citep{Miglio2013, Mathur2016}. In addition, with the automation in \pbjam we also aim to make asteroseismology accessible to non-specialists, thereby enabling its use in a wider range of contexts. The following refers to \pbjam v1.0.0\footnote{For the latest version see \url{https://github.com/grd349/PBjam}}.

\section{Peakbagging with \pbjam}
\begin{figure*}
    \centering
    \includegraphics[width = 2\columnwidth]{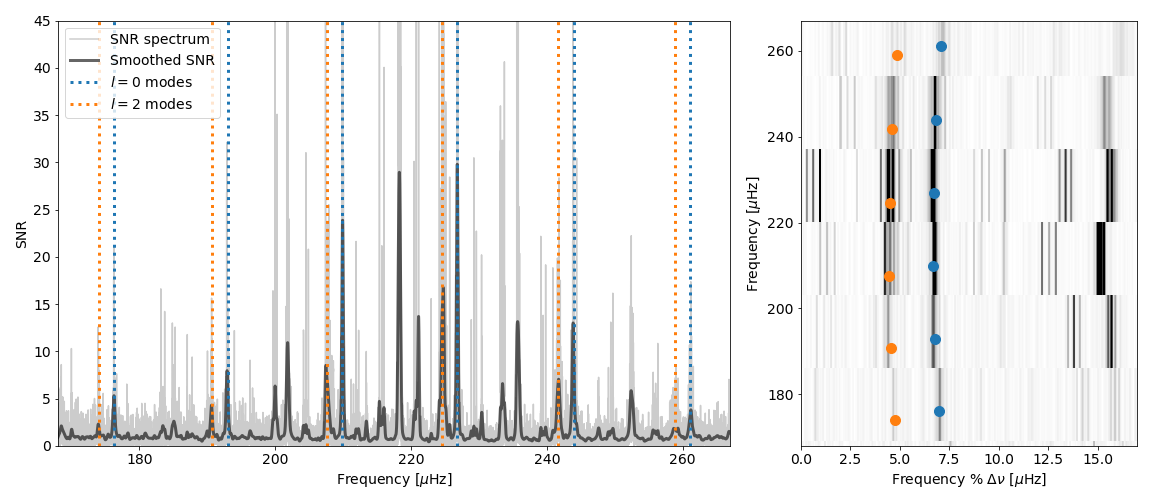}
    \caption{Left: SNR spectrum of the star \kicexample shown in light grey, and the smoothed spectrum in dark grey. The spectrum consists of a regular series of radial ($l=0$) and quadropole modes ($l=2$) modes shown by dashed lines. These are the modes that \pbjam attempts to model. The overall power in the \pmode envelope can be approximated by a Gaussian centered on $\numax$. Right: The spectrum shown as an échelle diagram, where the frequency axis is shown modulo the large separation, $\dnu=16.97\muHz$, showing the repeating mode pattern as distinct ridges. The points show the mode frequencies shown in the left frame, with the same color code.}
    \label{fig:example}
\end{figure*}
The purpose of peakbagging is the measurement of the oscillation mode frequencies of a star. An example spectrum is shown in Fig. \ref{fig:example}, where the modes of the RG star \kicexample\footnote{Additional examples can be found in the \pbjam GitHub repository.} are shown in terms of their SNR, which is the height of the mode peaks relative to the surrounding noise floor. The modes are centered around a characteristic frequency, $\numax$, and their SNR drops off rapidly toward higher and lower frequencies. This is usually called the \pmode envelope. 

The modes that are sufficiently excited to become visible vary depending on the physical properties of the star, as do the frequencies of the modes themselves. The number of visible radial orders depends on the SNR of the observations but typically range from just a few in the low SNR cases, to $\approx10-15$ in the best cases \citep{Lund2017, Chaplin2020}. The visibility of the modes also decreases with increasing $l$ \citep[see, e.g.,][]{Christensen-Dalsgaard1982}, and so usually only $l=0,1$ and $2$ are observed and rarely $l=3$ \citep{Appourchaux2012}. For each $l$ there are $2l+1$ modes with azimuthal order $-l \leq m \leq l$ which may become split in terms of frequency due to rotation of the star \citep{Gizon2003, Ballot2007}. The spectrum model therefore typically has a large number of variables, and some kind of parameterization of the model is usually necessary.

Several methods of parameterization have been used previously to study stellar power spectra \citep[see, e.g.,][]{Ballot2011a, Appourchaux2012, Handberg2011, Lund2017, Corsaro2019}. With this version of \pbjam the objective is to measure the mode frequencies of the $l=0,\,m=0$ and $l=2,\,m=0$ modes, which are highlighted in Fig. \ref{fig:example}. These modes contain information about the stellar properties for a large range of spectral types and evolutionary stages. The parameterization that is used in \pbjam is therefore greatly simplified compared to other studies, as effects such as rotation and mode asymmetry are ignored \citep[see, e.g.,][]{Davies2015, Benomar2018a, Benomar2018b}. Most significantly however, is that the $l=1$ modes are not included in the \pbjam peakbagging. These modes require special treatment to be fit accurately in an automated fashion across all evolutionary stages of a solar-like oscillator. This is particularly relevant for SG and RG stars, where \gmodes in the deep stellar interior start to couple with the $l=1$ \pmodes, resulting in a complicated pattern of mixed mode frequencies, which makes even manual mode identification challenging and an automated approach more so \citep[see, e.g,][]{Appourchaux2020}. We aim to include the $l=1$ treatment in future releases of \pbjam, as they carry useful information about rotation and help to further constrain stellar ages \citep{Metcalfe2010, Deheuvels2011}. However, the $l=0$ cannot couple to any gravity modes, and strong coupling of the $l=2$ $p$- and \gmodes is only rarely observed (Mosser et al. 2020, in prep), and so these mode pairs consistently appear with a near-regular pattern, at a rough interval called the large frequency separation, $\dnu$. This makes them far easier to identify in an automated fashion, and only using these mode frequencies is sufficient to determine the stellar mass and radius to a precision of a few percent, and to estimate the stellar age to within $\sim10-20\%$ \citep{Davies2016b, McKeever2019, Buldgen2019}. Figure \ref{fig:example} shows the spectrum of the star \kicexample where the $l=0,2$ pairs are highlighted. The repeating pattern appears clearly by showing the mode frequencies modulo the large separation, where the $l=0$ and $l=2$ modes align as ridges. Only some of the $l=1$ modes adhere to the same repeating pattern, while others are shifted due to the coupling with the internal gravity waves.

The choice of spectrum model and the mode identification determines the set of fit parameters, $\boldsymbol{\theta}$, and so to find the best-fit model we map the posterior probability
\begin{equation}
P(\boldsymbol{\theta}|D)\propto P(D|\boldsymbol{\theta})P(\boldsymbol{\theta}),    
\label{eq:posterior}
\end{equation}
where $P(\boldsymbol{\theta})$ is the constraint on the fit parameters given any prior knowledge, and $P(D|\boldsymbol{\theta})$ is the probability of observing the data $D$ given the model. If we use a more informal syntax then $P(\boldsymbol{\theta})$ is what we know already about what the mode frequencies should be, $P(D|\boldsymbol{\theta})$ is what the power spectrum or any other observations tell us about what the mode frequencies are. The posterior probability, $P(\boldsymbol{\theta}|D)$, is what we ultimately care about: the combination of our existing knowledge and our new knowledge.

\begin{figure}
    \centering
    \includegraphics[width = \columnwidth]{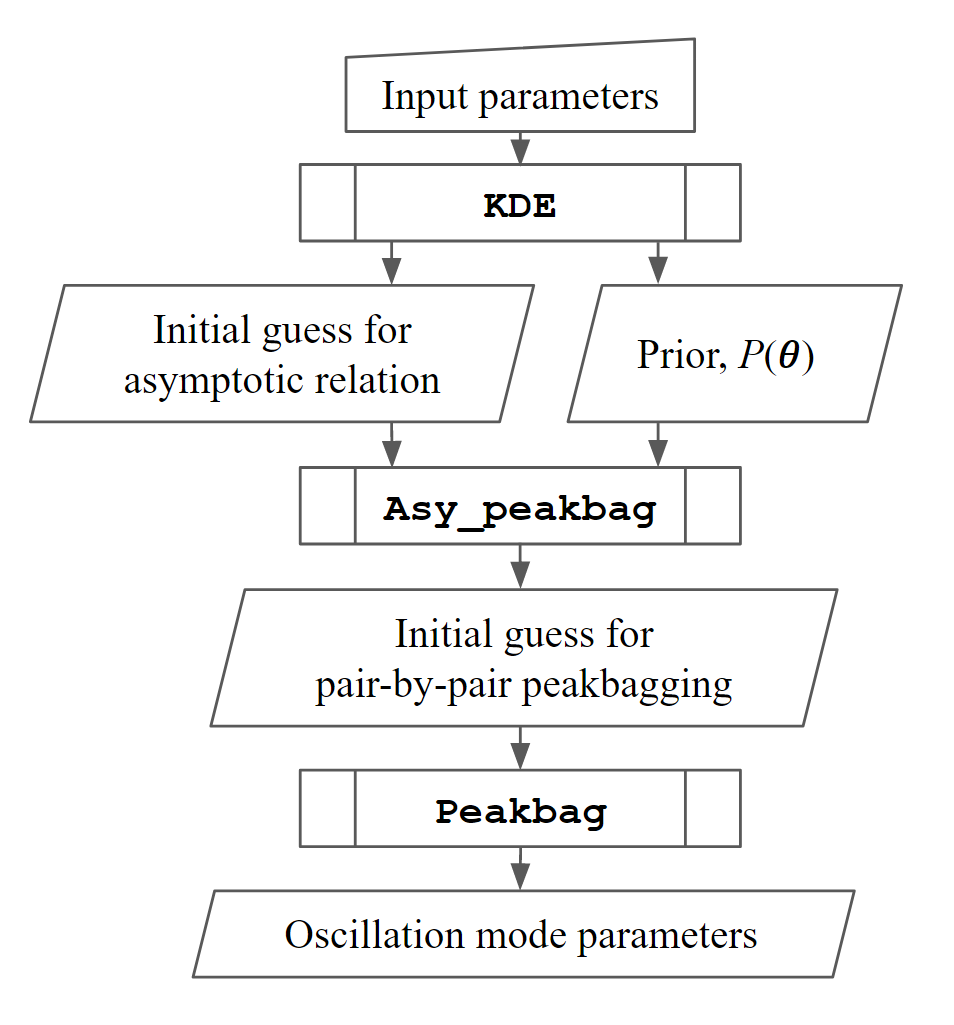}
    \caption{Diagram of the PBjam peakbagging process. The input parameters are shown in Table \ref{tab:asypars}. The output oscillation mode parameters include the mode frequencies, heights and widths along with their uncertainties.}
    \label{fig:flow}
\end{figure}

\section{Core \pbjam classes}
\label{sec:modules}

\begin{figure*}
    \centering
    \includegraphics[width = 2\columnwidth, trim={3cm 1cm 3cm 2cm}, clip]{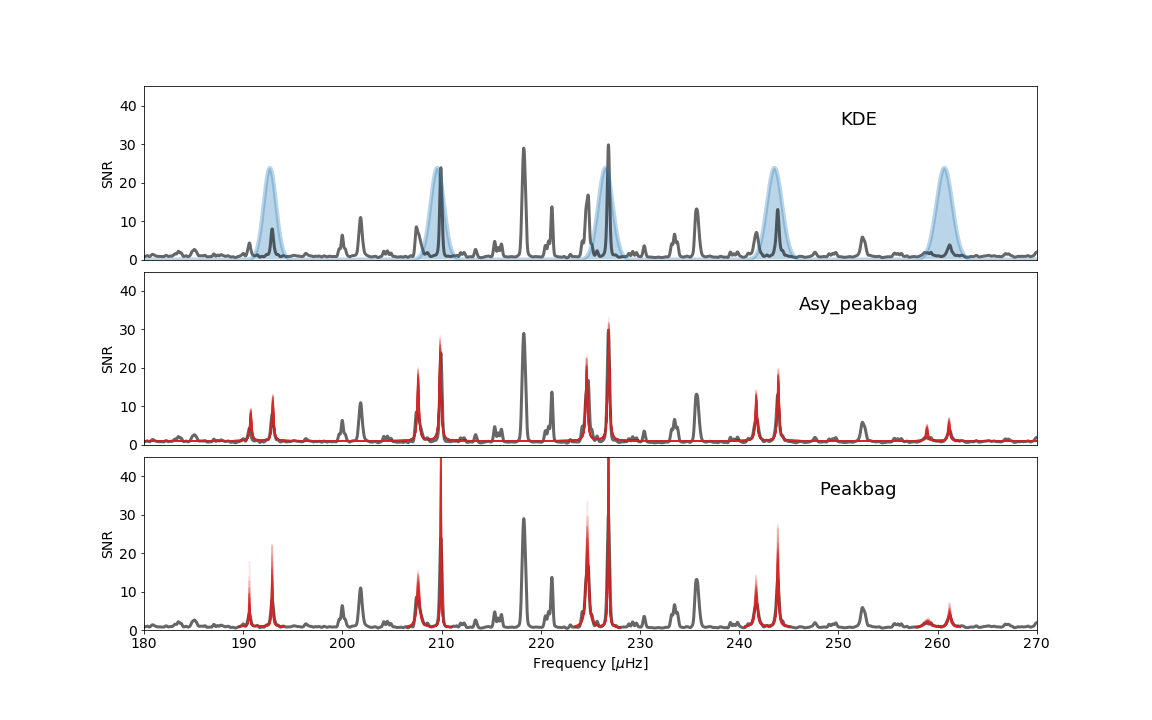}
    \caption{Outputs of the three main classes in \pbjam: \kde, \asypeakbag, and \peakbag. In all the frames the grey curve is the smoothed SNR spectrum of \kicexample. Top: the output of \kde is a probability distribution showing the location of the radial ($l=0$) modes, based on what the prior sample predicts. This stage is independent of the observed spectrum. Middle: The asymptotic relation is fit to the spectrum, using the probability distribution from \kde as input. A sample of 50 models drawn from the posterior distribution of the fit is shown in red. Bottom: The mode frequencies define the frequency ranges to be considered in the final peakbag step, where each $l=0,2$ pair is then fit independently. For each pair 50 models drawn from the posterior distributions are shown in red.}
    \label{fig:pbjamjoint}
\end{figure*}

The main functionality in \pbjam is the mode identification and model fitting, and it is currently comprised of three connected classes: \kde and \asypeakbag which perform the mode identification and encode the prior knowledge of the modes, and \peakbag which performs the final model fit but with minimal influence from the prior. Provided with a set of basic inputs shown in Table \ref{tab:asypars}, these classes are run in sequence to produce an estimate of the mode frequencies in the spectrum. The three steps can be executed automatically by \pbjam, but each may still be executed individually if needed. The automated process in \pbjam is illustrated in Fig. \ref{fig:flow}, with detailed descriptions in the following sections. The outputs of each step are summarized in Fig. \ref{fig:pbjamjoint}.
\begin{figure*}
    \centering
    \includegraphics[width = 2\columnwidth, trim={2.5cm 3cm 3cm 4cm}, clip]{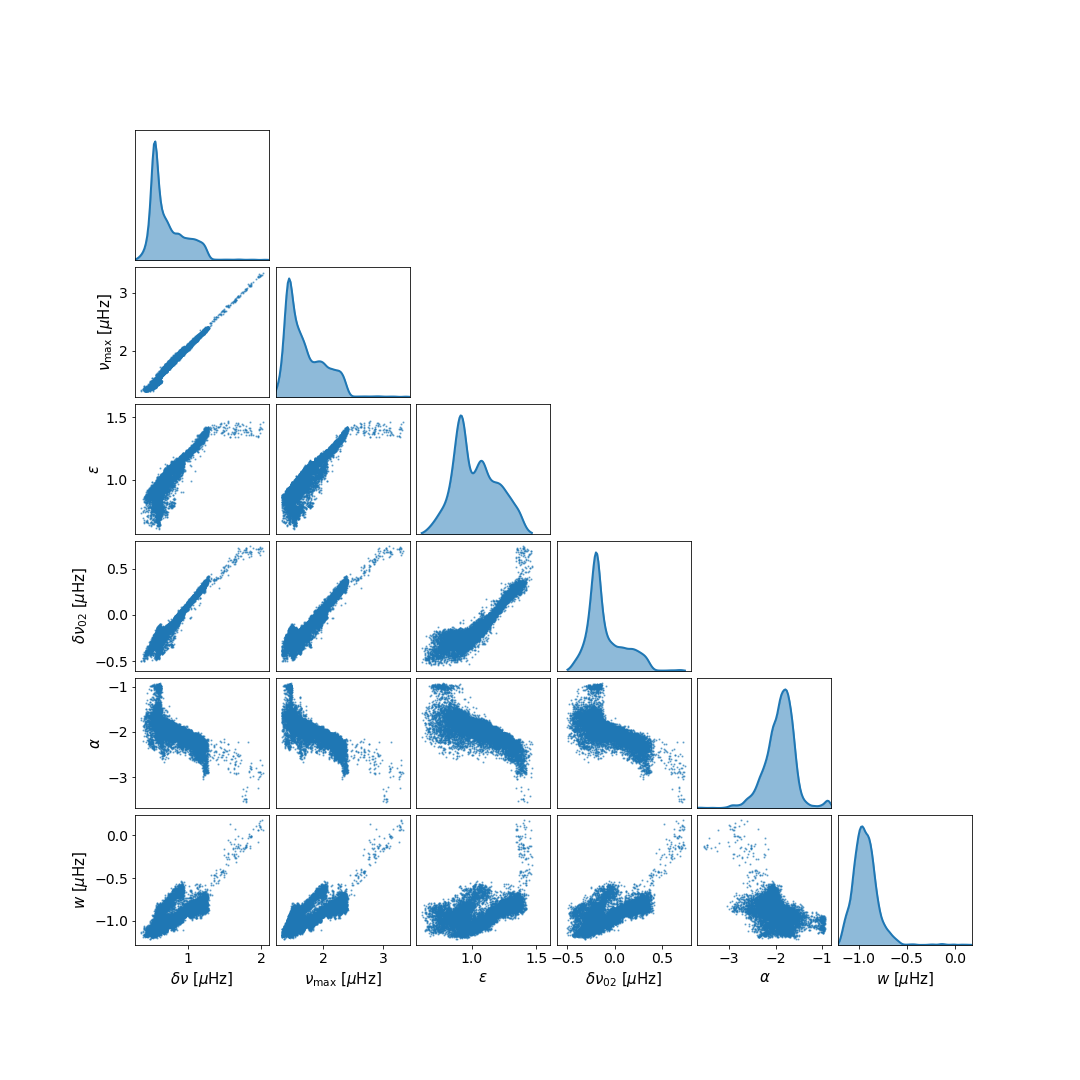}
    \caption{Corner plot of parameters in the \asypeakbag model. The diagonal shows the distribution of each model parameter, for stars in the current sample used by the \kde class to generate a prior. The off-diagonal frames show the correlations between the different parameters. All parameters except $\eps$ are on logarithmic scales. While all \priornumber stars in our sample are shown, for clarity we only show a subset of the parameters. For a full list and description of the parameters see Table \ref{tab:asypars}.}
    \label{fig:kdecorner}
\end{figure*}

\subsection{KDE - encoding prior knowledge}
\label{sec:kde}
This section discusses the details of \kde. The two main functions of \kde are to construct a function that approximates our prior knowledge, and to provide initial guesses for the fit parameters used in the following step in \asypeakbag (see Section \ref{sec:asypeakbag}). The parameters, $\boldsymbol{\theta}$, that \kde defines a prior for, are therefore dependent on which inputs \asypeakbag requires (see Table \ref{tab:asypars}). The methodology may, however, be applied to a variety of parameterisations of the spectrum model, and we will therefore keep the notation in this section more general.

\subsubsection{Using a kernel density estimate as a prior}
The first part of \kde is to construct a prior function that encodes our knowledge of how the mode frequencies behave at various stages of stellar evolution. 

This is done by computing a multivariate kernel density estimate (KDE) of $\boldsymbol{\theta}$, based on fits to previously observed targets from \kepler\footnote{Future versions of \pbjam will include \corot and \tess observations}. This sample is shown in Fig. \ref{fig:kdecorner} for a subset of the parameters used in \asypeakbag. Provided the sample of previous observations covers the physically meaningful range in parameter space, the resulting \kde captures the covariance of the different parameters in the fit. This yields a continuous function which approximates the prior, $P(\boldsymbol{\theta})$. 

Constructing a KDE is fundamentally a data smoothing problem. Here we use a multivariate KDE with a bandwidth for each of the fit parameters, $\boldsymbol{\theta}$, of the asymptotic relation and a few others (see Section \ref{sec:asypeakbag}). This gives full control over the degree of smoothing of the prior data. The KDE is constructed as
\begin{equation}
P(\boldsymbol{\theta}) \propto \frac{1}{K}\sum_{i=1}^{K}\textbf{Q}_\textbf{H}(\boldsymbol{\theta}-\boldsymbol{\theta}_i),   
\label{eq:prior}
\end{equation}
where $\boldsymbol{\theta}_i$ is the parameters of one of $K$ previously fit stars used to construct the prior. $\textbf{Q}_\textbf{H}$ is the bandwidth matrix for which we use a multivariate Gaussian covariance matrix with all off-axis elements set to zero (sometimes called a D-type kernel):
\begin{equation}
\textbf{Q}_\textbf{H}(\boldsymbol{\theta})=(2\pi)^{-\frac{d}{2}}\left|\textbf{H}\right|^{-\frac{1}{2}}\exp{-\frac{1}{2}\boldsymbol{\theta}^T\textbf{H}^{-1}\boldsymbol{\theta}},    
\end{equation}
where $d$ is the number of dimensions of the multivariate KDE. Here,  $\textbf{H}$ is a $d \times d$ diagonal matrix 
\begin{equation}
\textbf{H}=\mathrm{diag}(q_1^2,\ldots,q_d^2),   
\end{equation}
where $q_i$ is a scalar and typically referred to as the bandwidth of one of the model parameters. 

Computing the KDE is done using the \texttt{statsmodels}\footnote{\url{https://www.statsmodels.org/stable/index.html}} Python package \citep{Seabold2010}. This package also determines the optimal bandwidth for each fit parameter using a cross validated maximum likelihood approach. The cross validation is performed on a subset of the $\sim100$ nearest neighbours of the target star in terms of $\numax$. This subset is chosen from a range of up to $\pm20\sigma$, where $\sigma$ is the uncertainty on the input $\numax$. This approach creates a KDE where the optimal bandwidth varies inversely with the local density of prior data points, depending on the target star. The effect is to reduce the bandwidth in regions of the prior sample with many stars, while increasing the bandwidth in regions where we have few samples, i.e. little knowledge. The bandwidth is optimized every time a new fit is performed, and the prior is therefore expected to become more informative as we populate the different regions of the prior sample. We intend this to be an on-going process as new stars are continuously being observed and investigated. However, the bandwidth can be scaled by the user, which may be necessary where the prior sample is particularly sparse, for example at low $\numax$ ($\lesssim20\muHz$). 

An example of the resulting probability distributions of the $l=0$ modes determined by the \kde class are shown in the top frame of Fig. \ref{fig:pbjamjoint}.

\subsubsection{Estimating the most probable starting point}
The second function of \kde is to estimate the most probable region in parameter space for \asypeakbag to start. This is done by sampling $P(\boldsymbol{\theta}|D)$ shown in Eq. \ref{eq:posterior}. The prior, $P(\boldsymbol{\theta})$, is approximated using a KDE as shown above, which leaves the likelihood $P(D|\boldsymbol{\theta})$. At this stage we use the input parameters, shown in Table \ref{tab:asypars}, as the observational data, $D$. These input parameters are: the frequency of maximum power of the \pmode envelope, $\numax$, the frequency difference of consecutive overtones, $\dnu$, the effective temperature of the star, $\Teff$ and the \emph{Gaia} photometric color index $\bprp$ \citep{Evans2018}. The likelihood is the joint probability estimated by a series of normal distributions given by these input parameters and their uncertainties. The uncertainties on the inputs are thereby accounted for when estimating the initial starting location. Since $\Teff$ and $\bprp$ contain much of the same information for characterizing a star, only one of these parameters need be provided. In the event that only one of these parameters is available, \pbjam assumes a very wide, uninformative prior on the missing parameter. 

Sampling the posterior, $P(\boldsymbol{\theta}|D)$, is done using the affine-invariant MCMC sampler from the \emcee\footnote{\url{https://emcee.readthedocs.io/en/stable/}} \citep{Foreman-Mackey2013} package. The resulting percentile values of the marginalized posterior of each parameter are then passed to \asypeakbag as the initial starting location in parameter space.

\subsection{Asy\_peakbag - Initial guesses from the asymptotic relation}
\label{sec:asypeakbag}
This section discusses the details of \asypeakbag, which fits the asymptotic relation \citep[see, e.g.,][]{Mosser2015} to the spectrum, using the output of the \kde class. This in turn provides the most probable frequency intervals for the final stage of peakbagging (see Section \ref{sec:peakbag}).

\asypeakbag essentially performs a highly constrained peakbagging fit, which may in principle be used to estimate the mode frequencies on its own. However, the asymptotic relation used here is not a complete description of the mode frequencies, and so will not capture detailed variations related to, e.g., acoustic glitches \citep{Mazumdar2014, Vrard2019} and coupling to \gmodes in the deep interior of the star \citep{Mosser2017}. Furthermore, any error estimates on the asymptotic fit parameters will be highly correlated. \asypeakbag is therefore best used to simply define a credible frequency range of the individual modes, and to identify the angular degree, for more detailed peakbagging later (for example using the \peakbag class). 

In this case the spectrum model, $M(\boldsymbol{\theta}, \nu)$, consists of a sum of Lorentzian profiles, one for each visible mode in the spectrum \citep{Anderson1990}. For the purposes of \pbjam we treat the $(n,l=0)$ and $(n-1, l=2)$ as a pair, where $n$ represents the visible radial orders in the spectrum.

The model for \asypeakbag is then
\begin{equation}
    M\left(\boldsymbol{\theta},\nu\right)=b+\sum\limits_{n=1}^{N}\frac{h_{n,0}}{1+\frac{4}{w^2}(\nu-\nu_{n,0})^2}  +\frac{h_{n-1,2}}{1+\frac{4}{w^2}(\nu-\nu_{n-1,2})^2}.
    \label{eq:asymod}
\end{equation}

We fit Eq. \ref{eq:asymod} to the SNR spectrum of a star and we therefore assume that the constant term $b=1$. The SNR spectrum is the power spectrum divided by the background noise level, which is approximated by a running median of the power spectrum\footnote{See \url{https://docs.lightkurve.org/api/lightkurve.periodogram.Periodogram.html}}. The precise treatment of the background noise level caused by granulation and long period variability may influence the measurement of the mode frequencies. However, the method applied here is sufficient for the purposes of setting credible frequency ranges for later detailed peakbagging. 

Rather than fitting a series of free parameters for each mode in Eq. \ref{eq:asymod}, we use the asymptotic relation to encode our expectation of the mode pattern in solar-like oscillators. The mode frequencies are then given by
\begin{equation}
\begin{array}{ll}
\nu_{n,0}  & =\left(n+\eps+\frac{\alpha}{2}\left(n-n_{\mathrm{max}}\right)^2\right)\dnu\\
\nu_{n-1,2} & =\nu_{n,0}-\dotwo,
\end{array}
\label{eq:asymptotic}
\end{equation}
where $\eps$ is commonly referred to as a frequency offset, or phase term, and $n_{\mathrm{max}}=\numax/\dnu-\eps$ \citep[e.g.,][]{Kjeldsen2005}. The parameter, $\alpha$, is the scale of the second order variation of the modes frequencies between each radial order. The frequency of the quadrupole, $l=2$, modes are offset from the radial modes by $\dotwo$, which is assumed constant for all the radial orders in the fit. 

The relative mode heights, in terms of SNR, are approximated by a Gaussian envelope,
\begin{equation}
\begin{array}{ll}
    h_{n,0} & = H_{\mathrm{max}}\exp\left(-0.5\left(\nu-\numax\right)^2 / W^2_{\mathrm{env}}\right), \\
    h_{n-1,2} & = 0.7\,h_{n,0},
\end{array}
\end{equation}
where $W_{\mathrm{env}}$ is the \pmode envelope width, and $H_{\mathrm{max}}$ is the envelope height. The heights of the $l=2$ modes are scaled by a factor of 0.7, relative to those of the $l=0$ modes \citep{Ballot2011b}. This approximates the reduced visibility due to geometric cancellation as the degree $l$ of the modes increases.

The widths of the Lorentzian profiles are known to be a function of $\Teff$, $\numax$, and mode frequency \citep[see, e.g.,][]{Appourchaux2014}. However, precisely modeling these requires several additional parameters. This adds complexity to the model, which for the purposes of \asypeakbag is unnecessary, and we therefore simply approximate the mode widths by a single constant value of $w$ for all modes.

With \asypeakbag we also evaluate the posterior shown in equation Eq. \ref{eq:posterior}, however here we add the additional constraints from the power spectrum, i.e., the model fit.

This means that the logarithm of the posterior probability can be written as
\begin{equation}
\ln{P(\boldsymbol{\theta}|D)}\propto \ln{\mathcal{L}(\boldsymbol{\theta})} + \ln{P(\boldsymbol{\theta})},
\end{equation}
where  $P(\boldsymbol{\theta})$ is given by Eq. \ref{eq:prior} and $\mathcal{L}(\boldsymbol{\theta})$ is the observational constraints given by
\begin{equation}
    \ln{\mathcal{L}(\boldsymbol{\theta})} = \ln{\mathcal{L}_{\mathrm{S}}(\boldsymbol{\theta})} + \ln{\mathcal{L}_{\mathrm{O}}(\boldsymbol{\theta})}. 
\end{equation}

The log-likelihood $\ln{\mathcal{L}_{\mathrm{S}}(\boldsymbol{\theta})}$ is the constraint from the power spectrum. The power in a frequency bin of the power spectrum is Gamma distributed with a shape parameter $\alpha=1$ and scale parameter $\beta=1/M(\boldsymbol{\theta},\nu)$, and so the log-likelihood may be computed by \citep[see, e.g.,][]{Woodard1984, Duvall1986}
\begin{equation}
 \ln{\mathcal{L}_{\mathrm{S}}\left(\boldsymbol{\theta} \right)} = - \sum\limits_{j=1}^J {\left(\ln M\left( {\boldsymbol{\theta} ,\nu _j } \right) + \frac{{S_j }}{{M\left( {\boldsymbol{\theta} ,\nu _j } \right)}}\right)},
\end{equation}
where $S_j$ is the power in frequency bin $j$, and $J$ is the total number of frequency bins in the spectrum.

Here, the log-likelihood $\ln{\mathcal{L}_{\mathrm{O}}(\boldsymbol{\theta})}$ is from the additional observational parameters where, similarly to \kde, \asypeakbag also uses $\Teff$ and $\bprp$. However, $\numax$ and $\dnu$ are now only used as fit parameters, with the only constraint on these parameters coming from the prior and the spectrum, and not from the user input. This prevents double-counting the information from the input $\numax$ and $\dnu$ and the spectrum itself, as the former is often derived from the same spectrum that is being fit. The constraints from $\Teff$ and $\bprp$, are again simply a sum of normal distributions with mean values and standard deviations equal to the inputs provided by the user.

The KDE that represents the prior, $P(\boldsymbol{\theta})$, is constructed from a sample of \priornumber \kepler stars ranging from RG stars with a $\numax \approx 30\muHz$ to MS stars at $\numax\approx4000\muHz$ (see Fig. \ref{fig:kdecorner}) that were previously fit using Eq. \ref{eq:asymod}. This sample\footnote{Available in machine readable format \url{https://github.com/grd349/PBjam/blob/master/pbjam/data/prior_data.csv}.} is compiled from the \citet{White2011}, \citet{SilvaAguirre2015}, \citet{Serenelli2017}, \citet{Lund2017}, and  \citet{Yu2018} catalogs, with the intent that more are added as they become available.

To sample the parameter space spanned by $\boldsymbol{\theta}$, \pbjam can use either the \emcee or \cpnest \footnote{\url{https://johnveitch.github.io/cpnest/}} \citep{Veitch2017} packages, either of which provide an estimate of the posterior probability distribution $P(\boldsymbol{\theta}|D)$. The marginalized posterior distributions of the fit parameters are then passed to the next step, \peakbag.

An example of fitting the asymptotic relation to the spectrum of \kicexample can be seen in the middle frame of Fig. \ref{fig:pbjamjoint}.

\begin{table}
    \centering
    \caption{Fit parameters in \kde and \asypeakbag. Parameters marked with $\dagger$ are required input values which must be computed or estimated by the user, in order to start \kde. Variables marked with $\ast$ are only used to remove degeneracies in the asymptotic relation, and are not part of the spectrum model itself.}
    \begin{tabular}{lp{54mm}}
        Parameter & Description \\
        \hline
        $\Teff$$^{\ast\dagger}$ & Effective surface temperature of the star.\\
        $\bprp$$^{\ast\dagger}$ & \emph{Gaia} photometric color index. \\
        $\dnu$$^{\dagger}$ & Frequency difference of modes with the same $l$, but consecutive $n$.\\
        $\numax$$^{\dagger}$ & Frequency of maximum power of the \pmode envelope.\\
        $\eps$ & Radial order phase term.\\
        $\dotwo$ & Frequency difference of $l=0$ and $l=2$ modes.\\
        $\alpha$ & Scale (curvature) of radial order variation with frequency.\\
        $H_{\mathrm{max}}$ & Height of the \pmode envelope.\\
        $W_{\mathrm{env}}$ & Width in frequency of the \pmode envelope.\\
        $w$ & Width in frequency of the modes. 
    \end{tabular}
    \label{tab:asypars}
\end{table}

\subsection{Peakbag - pair-by-pair mode fitting}
\label{sec:peakbag}
The final part of the process is to relax the majority of the parametrization used in \asypeakbag, thereby providing a relatively unconstrained estimate of the mode frequencies and their uncertainties. This allows \pbjam to capture small variations in the mode frequencies, that are not accounted for by the asymptotic relation. 

We start our description of the \peakbag spectrum model by once again only considering the $l=0,2$ pairs as is done in with \asypeakbag, and we can therefore use Eq. \ref{eq:asymod}. However, in the \peakbag model we only consider the frequency range near the $l=0,2$ pairs, and not the range in between the pairs that is typically occupied by the $l=1$ modes. This is done to speed up this final peakbagging step. In the following, the $l=0,2$ pairs are therefore treated independently, and for simplicity we will simply denote each $(n, l=0)$, $(n-1, l=2)$ pair as $(n, 0),$ $(n, 2)$, so that $n$ is now the pair number and not the radial order. We then have a model, $M_n(\nu)$, for each mode pair and a small segment of the spectrum surrounding them (see bottom frame of Fig. \ref{fig:pbjamjoint}).

In \peakbag we use \pymc \citep{Salvatier2016} to sample the posterior distribution. \pymc is a probabilistic programming language and so we will adopt consistent notation here. This allows us to set up random variables to describe the model parameters.  The mode frequencies of a pair $n$ are then given by
\begin{equation}
\nu_{n,l} \sim \mathcal{N}(\nu^{\mathrm{asy}}_{n,l},\left(0.03\dnu\right)^2) \quad \mathrm{for}\quad l=0,2,
\label{eq:Nnu}
\end{equation}
which denotes $\nu_{n,0}$ and $\nu_{n,2}$ as random variables that are distributed according to a normal distribution with mean mode frequencies, $\nu^{\mathrm{asy}}_{n,l}$, from \asypeakbag, and variances of $(0.03\dnu)^2$. 

If we were to turn Eq. \ref{eq:Nnu} into a \pymc model and draw samples from it, we would get posterior distributions that are normally distributed with mean values equivalent to the \asypeakbag predictions and standard deviations that are $3\%$ of $\dnu$. 

We use the same arrangement for mode heights and widths. However, these parameters can only take on positive values, and so to encode this knowledge we use log-normal distributions instead. As the prior mean values we again use the estimates from \asypeakbag. However, for the mode heights we use standard deviation of $0.4$ dex, and $1$ dex for the mode line widths. The large range for the latter is required since the line widths vary widely with frequency across the \pmode envelope, and may be much smaller or larger than the approximate mean value estimated by \asypeakbag. 

The mode heights are then
\begin{equation}
h_{n,l} \sim \log{\mathcal{N}\left(\log{h^{\mathrm{asy}}_{n,l}}, 0.16\right)} \quad \mathrm{for}\quad l=0,2.\\  
\end{equation}

Similarly for the line widths 
\begin{equation}
w_{n,l} \sim \log{\mathcal{N}\left(\log{w^{\mathrm{asy}}},1.0\right)}\quad \mathrm{for}\quad l=0,2 .   
\end{equation}

We construct a prior for the background $b_{n}$ in a similar fashion. As above, we are working with the SNR spectrum and so the background should be unity. However, the flattening becomes an increasingly bad approximation as the SNR ratio of the modes increases. This is typically a small effect, but to provide a more robust result we allow the background term for each pair of modes to vary independently, while still being consistent with the same log-normal distribution. We therefore set up the background prior as
\begin{equation}
b_n \sim \log{\mathcal{N}\left(0, 0.16\right)}.    
\end{equation}

The final step of building our \pymc model is adding the constraint from the spectrum itself. As with \asypeakbag we assume that the power in each bin is drawn from a Gamma distribution, however, here the scale parameter $\beta=1/M_n(\nu)$, which defines a model for each mode pair $n$. In \pymc terminology the spectrum constraint is then imposed by
\begin{equation}
S_n \sim \gamma(1,1/M_{n}(\nu)).    
\end{equation}

This completes our \pymc model, and we use the No U-turn Sampler (NUTS) to draw samples from the posterior probability distribution. An example of the models sampled by \peakbag is shown in the bottom frame of Fig. \ref{fig:pbjamjoint}. The output is in the form of summary statistics of each parameter posterior distribution. As a guide to the user, one of the output metrics is the ratio of the prior (Eq. \ref{eq:Nnu}) and the posterior widths of the mode frequencies. For modes where this ratio is $\sim 1$ the inference is influenced mainly by the prior, and likewise when the ratio is $>1$ the spectrum provides the most information. The former often being the case for low SNR modes, and the latter for high SNR modes. While \pbjam does not make any selection based on this metric, we suggest that modes with prior to posterior width ratios greater than $\approx2$ can be used to select modes where the spectrum dominates the inference. However, even when the inference is informed predominantly by the prior, the resulting mode frequency may still be used as a constraint on stellar models. It then simply reflects the estimate gained from the asymptotic relation shown in Eq. \ref{eq:asymptotic}, but with larger uncertainties than if the spectrum also contributes to the inference.

\section{Conclusion}
\label{sec:conclusion}

The current version of \pbjam is suitable for measuring the $l=0$ and $l=2$ mode frequencies from main-sequence stars up to red-giant stars. This is done by using the expected pattern of the mode frequencies, provided by the asymptotic relation and the wealth of asteroseismic data provided by space missions such as \kepler. Apart from providing a few input parameters, the mode frequencies are peakbagged in a completely automated fashion. This allows non-expert users to obtain seismic constraints for a wide range of problems that involve solar-like oscillators. 

This sample of prior targets used by \pbjam contains \priornumber stars observed by \kepler. As shown in Fig. \ref{fig:kdecorner} this sample currently contains predominantly red giant stars, and only few main-sequence stars. This is due to telemetry restrictions on the \kepler spacecraft, which limited the number of targets that could be observed in high cadence mode. Main-sequence stars typically have oscillation frequencies $\gtrsim1000\muHz$, and so oscillations were only detected in a few hundred of these stars. It is our goal to keep updating the sample of observed targets as they are observed and peakbagged. The \tess mission is promising to contribute substantially in the respect, as the observing strategy covers almost the entire sky, and includes many bright sub-giant and main-sequence stars.

The main limitation of the current version of \pbjam is that it does not consider the $l=1$ modes. While, the $l=0,2$ mode pairs alone are sufficient to constrain stellar models to a precision of a few percent in mass and radius, and $10-20\%$ in age, the $l=1$ mode frequencies will provide much tighter constraints. The coupling of the $l=1$ \pmodes to the internal gravity dominated oscillations and how these change as the star evolves, is the main issue with fitting these modes. Previous work by \citet{Kallinger2019}, \citet{Corsaro2020}, \citet{Appourchaux2020} among others, have sought to address this issue in an automated fashion. These methods tend to use mode-by-mode significance tests of the peaks in the spectrum, followed by a comparison with the asymptotic relation to identify both the $l=0,2$ pairs and subsequently the $l=1$ modes.

The aim of \pbjam is to create a generative model that leverages the information from the whole \pmode envelope at once, and which scales evenly from the main-sequence up to the red giant branch. This has largely been achieved for the $l=0,2$ pairs in the current version, by first establishing the prior frequency range from the asymptotic relation, and then subsequently releasing those constraints. However, applying this approach for fitting the $l=1$ modes requires that the knowledge of the expected mode pattern is accurate. While this is comparatively straightforward for main-sequence stars, and to an extent red giant stars \citep{Vrard2016}, it is less so for sub-giants. For these stars even small errors in the $l=1$ peakbagging can have large effects on the resulting stellar parameters \citep[see, e.g,][]{Li2020}. Extending the application of prior information, as is done in \pbjam, to potentially incorporate the approaches mentioned above is one possible solution. We leave this investigation to the next release of \pbjam.

\software{astropy \citep{astropy2018},  
          corner \citep{Foreman-Mackey2016},
          CPNest \citep{Veitch2017}
          emcee \citep{Foreman-Mackey2013},
          matplotlib \citep{Hunter2007},
          numpy \citep{Oliphant2006},
          lightkurve \citep{Barentsen2019},
          pandas \citep{Reback2020},
          Python \citep{python1995}
          pymc3 \citep{Salvatier2016},
          scipy \citep{Virtanen2020},
          sklearn \citep{Pedregosa2011},
          statsmodels \citep{Seabold2010}
        }

\acknowledgments
MBN, WHB, and WJC acknowledge support from the UK Space Agency. GRD, OJH and WJC acknowledge the support of the UK Science and Technology Facilities Council (STFC). EC acknowledges support from PLATO ASI-INAF agreement n.2015-019-R.1-2018 and by the European Union’s Horizon 2020 research and innovation program under the Marie Sklodowska-Curie grant agreement No. 664931. P. Gaulme acknowledges funding from the German Aerospace Center (Deutsches Zentrum für Luft- und Raumfahrt) under PLATO Data Center grant 50OO1501. JO acknowledges support from TESS GI grant G022092. RAG acknowledges support from the GOLF and PLATO/CNES grants. This paper has received funding from the European Research Council (ERC) under the European Union’s Horizon 2020 research and innovation programme (CartographY GA. 804752). The authors acknowledge use of the BlueBEAR HPC service at the University of Birmingham. Funding for the Stellar Astrophysics Centre is provided by The Danish National Research Foundation (Grant agreement no.: DNRF106). This paper includes data collected by the Kepler mission and obtained from the MAST data archive at the Space Telescope Science Institute (STScI). Funding for the Kepler mission is provided by the NASA Science Mission Directorate. STScI is operated by the Association of Universities for Research in Astronomy, Inc., under NASA contract NAS 5–26555. This work has made use of data from the European Space Agency (ESA) mission {\it Gaia} (\url{https://www.cosmos.esa.int/gaia}), processed by the {\it Gaia} Data Processing and Analysis Consortium (DPAC, \url{https://www.cosmos.esa.int/web/gaia/dpac/consortium}). Funding for the DPAC has been provided by national institutions, in particular the institutions participating in the {\it Gaia} Multilateral Agreement.

\bibliography{main}{}
\bibliographystyle{aasjournal}

\end{document}